# Theoretical introduction to express way of the complex determination of the thermophysical parameters of metals

Yuri Kornyushin

*Maître Jean Brunschvig Research Unit, Chalet Shalva, Randogne, 3975-CH*

A method to measure thermophysical characteristics (TPC) of conductive materials is described. The method is based upon controlled bulk (e.g., Ohmic) heating and measurement of the averaged over the volume of a sample temperature and the temperature of the surface of a sample. To measure averaged over a sample volume temperature a dilatometric technique could be used and surface temperature could be measured by x-ray method. This allows one to avoid distortion of the temperature field by the thermocouples and directly takes into account heat losses from the surface of a sample. The treatment of the experimental data should be based upon theoretical calculations presented in this paper. The measured values of the averaged over the volume of a sample temperature and a surface temperature allow for the calculation of TPC of the material. The experimental data obtained on the iron-based materials prove the high accuracy of the described approach [1-3].

## 1. Introduction

X-ray diffraction and dilatometric methods are often used for the purpose of investigation of the thermal properties of condensed matter [4]. Information on the temperature dependence of TPC of the construction materials is needed to create products of modern engineering and operate them in a wide temperature range. To determine TPC of a material the inverse problem of the thermal conductivity theory is to be solved. The coefficients of the thermal conductivity equation are usually found with the aid of the temperature field and boundary conditions, which are known from the experiment. That is why measurements (or setting) of the temperature (or thermal fluxes) in a sample in two points at least are usually a basis for the method of determination of TPC. Hence, accuracy of the measurements is determined by both the precision of the means of measuring as well as the quantity of the measured differences. Moreover, the accuracy of the measurements depends essentially on a series of other parameters: geometry and intrinsic heat capacity of the temperature transducers, method of realization of heat contact, accuracy of the measurements of the transducer coordinates, taking into account the heat losses, etc. That is why determination of TPC at elevated and high temperatures is a problem of special difficulty [5]. Errors are as large as several tens of centigrades when thermocouples are used as transducers for measurements on metals [6].

Values of the coefficients of the thermal conductivity $\lambda$ are high for metals and this is an obstacle for application of wide spread techniques of radial heat flux [7-9]. The measurement error in this case is about 30% as estimated in [7]. To create the temperature difference, which can be measured reliably, methods of the longitudinal heat flux are applied [10-13]. Measurements of heat fluxes in calorimetric cameras of diverse designs are used also [5,14,15]. Designs, which allow one to perform combined measurements, are of special interest [16]. However, high accuracy calorimeters are rather expensive and low efficient.

In 1987 Meir E. Gurevich in fond memory had proposed a method of combined determination of TPC of metallic materials [1-3]. This method allows for more efficient determination of TPC with higher precision; it also expands the area of application to nondestructive methods of examination of the temperature fields in metals, half-finished products and wares. The author of this paper had performed necessary for realization of this method theoretical calculation. This calculation is the subject of present paper.

## 2. Parametric measurements of the temperature of a sample

There is a possibility to determine the temperature of a sample using measurements of the dimension of a body as a function of its temperature:

$$\delta T = (1/\alpha)\delta h/h_0, \qquad (1)$$

where $\delta T = T - T_0$ is the change in the body temperature $T$ ($T_0$ is a starting temperature), $\delta h = h - h_0$ is the change in the linear size of a body $h$ ($h_0$ is a starting linear size), $\alpha$ is the linear coefficient of the thermal expansion.

When the temperature of the body is the function of coordinates, Eq. (1) determines the temperature averaged over the sample volume. Measurement of the dilatation of the homogeneously heated body in a thermostatic medium had been used in [17] to determine heat and temperature conductivity coefficients in silver and aluminum. Different types of linear and bulk dilatometers can be used for the dilatometric measurements [18]. A method of x-ray dilatometry was used in [1,3].

Changes in the lattice parameter, registered by x-ray diffraction technique during heating, can be surely referred to the surface layer, which is responsible for the heat exchange with the environmental medium. So these measurements correspond to the thermal dilatation, caused by the temperature of the surface of a sample:

$$\delta T_s = \delta b/\alpha b, \qquad (2)$$

where $T_s$ is the temperature of the surface, $\delta T_s = T_s - T_0$ is the change in the temperature of the surface of a body, and $\delta b = b - b_0$ is the change in the lattice parameter $b$ during heating ($b_0$ is the initial lattice parameter).

So the difference in the temperature in a sample can be determined using methods which do not distort the field of the temperature and then, taking the boundary conditions into account, one can calculate the heat transfer coefficients.

According to the definition the specific (per unit volume) heat capacity is

$$c = dQ/dT = q/T_t, \qquad (3)$$

where $Q$ is the enthalpy (per unit volume) of a sample, $q$ is the specific (per unit volume) power, spent to change the temperature of a sample during its heating with the rate $T_t = \partial T/\partial t$.

Hence, to determine veritable heat capacity of a material one ought to maintain adiabatic conditions measuring $q$, or to measure and take into account the capacity of the heat losses from the surface of a sample. The rate of temperature change during measurements in adiabatic calorimeters is from 1.0 to 10 K/min. In a sample heated by electric current there is a difference in the temperature in the center of a sample and its periphery. This difference arises due to the heat losses from the surface, heat inertia, and the inertia of the automatic system of following of adiabatic curve. That is why during calculation of $c(T)$, measured quantity of heat or heat power ought to be referred to the maximal temperature, obviously achieved in the center of a sample, in a case of a bulk heating.

On the other hand, conditions of heating can be regarded as nearly adiabatic when rate of the input of heat into the sample is essentially larger than the rate of the heat output. Such conditions are realized approximately at high rate heating.

Measurements aimed to determine the temperature of the surface of a sample should take into account heat losses and this leads to the boundary value problem of the first type in the heat conductivity theory.

Relationships for TPC, taking into account heat losses from the surface of a sample, heated by electric current, will be presented below.

## 3. Basic relations to calculate TPC

In case of low frequency current (the standard frequencies are 50 and 60 Hz) Ohmic heating can be considered as uniformly spread over the section of the conductor, i.e. one can regard homogeneously distributed internal sources of heat with a specific power $q$ in these cases.

The dependence of the temperature of the body $T$ on coordinates $x$, $y$, $z$, and time $t$ is determined by the heat conductivity equation [19]:

$$\lambda LT + q = cT_t, \tag{4}$$

where $L = (\partial^2/\partial x^2) + (\partial^2/\partial y^2) + (\partial^2/\partial z^2)$ is the Laplace operator.

Taking into account boundary conditions is the most complicated part of the heat conductivity problems. Newton convective exchange conditions or (and) radiation conditions lead to overcomplicated solutions of the heat conductivity problems, which make the physical sense of described phenomena obscure. The situation demands regarded problems to be simplified. Calculating temperature fields we shall neglect the thermal expansion of a material. Another simplification is taking into account that there is a possibility to measure simultaneously several quantities, which determine a problem, and to use them as parameters of a description.

Solving Eq. (4) we consider the temperature on the surface of a body $T_s$ being a given function of coordinates and time. Let us restrict ourselves to consideration of the homogeneous samples of small sizes. Small samples are convenient to measure TPC of materials. Time required for the heat exchange in such samples is rather small.

Let the temperature be represented as

$$T(\mathbf{r},t) = \langle T \rangle + \tau(\mathbf{r},t), \tag{5}$$

where $\langle T \rangle$ is the temperature averaged over the volume of a body, and $\tau(\mathbf{r},t)$ is much smaller than $\langle T \rangle$.

Such representation is justified in case of the uniform heating under conditions when inhomogeneity of the temperature due to heat losses has not enough time to become essential. Bearing in mind Eq. (5), let us restrict ourselves by regarding only linear terms in respect to the space and time derivatives of $T$. In such an approximation $c$, $q$, and $\lambda$ ought to be regarded as functions of $\langle T \rangle$ or $T_s$ in Eq. (3). Under usual conditions of experiment (homogeneous samples in which there are no areas where $c$, $\lambda$, and specific electric conductivity $\sigma$ differ by the orders of magnitude), when the time of heat exchange in a sample $R^2c/\lambda$ ($R$ is the characteristic size of a sample) is not large enough for $\langle T \rangle - T_s$ to change essentially, inequality

$$R^2 \ll \lambda/c \left| \partial \ln[(\langle T \rangle - T_s)/T_0] \partial t \right| \tag{6}$$

is fulfilled, which is the criterion for the quasi-stationary condition. In this case $c\partial\tau/\partial t$ can be neglected in respect to $\lambda L\tau$ and thus we have from Eq. (4):

$$L\tau = (c/\lambda)(\partial\langle T\rangle/\partial t) - (q/\lambda) \tag{7}$$

in linear on $\tau$ approximation.

Let us derive Eq. (6). The change in $(\langle T \rangle - T_s)$ during heat exchange time $(R^2c/\lambda)$ is as follows: $[\partial(\langle T \rangle - T_s)/\partial t](R^2c/\lambda)$. It should be much smaller than $(\langle T \rangle - T_s)$. From this follows Eq. (6). In Eq. (6) $T_0$ is irrelevant. It is introduced to have dimensionless quantity under ln function only.

Let us consider solutions of Eq. (7) for the cases of the thin slab, cylinder, and sphere. Samples of such shape are the most frequent in experiment.

## 3.1. Thin slab

Let us consider a thin slab of the thickness $2h$. Let $\langle T \rangle_h$ denotes the temperature averaged over the slab volume (coordinate $z$). Let $T_s$ can change smoothly along the surface of a sample, i.e. $T_s$ changes inessentially on the length $h$:

$$h \, |\partial \ln(T_s/T_0)\partial x| \ll 1, \; h \, |\partial \ln(T_s/T_0)\partial y| \ll 1. \tag{8}$$

Eq. (5) for a slab is

$$T(x,y,z,t) = \langle T \rangle_h + \tau(x,y,z,t), \; \tau(x,y,z,t) \ll \langle T \rangle_h. \tag{9}$$

Temperature $\langle T \rangle_h$ depends on time and can vary smoothly along the slab surface. Due to the smoothness of the dependence of quantities on $x$ and $y$ it is possible to neglect $(\partial^2\tau/\partial x^2) + (\partial^2\tau/\partial y^2) = L_{xy}\tau$ comparatively to $\partial^2\tau/\partial z^2 = L_z\tau$ in the thermal conductivity equation and to obtain from Eq. (7) following relationship:

$$L_z\tau = (c/\lambda)\langle T_t \rangle_h - (q/\lambda) - L_{xy}\langle T \rangle_h. \tag{10}$$

Here $T_t = \partial T/\partial t$. In Eq. (10) $x$, $y$, and $t$ are just parameters, not variables.
As follows from Eq. (5) boundary condition to Eq. (10) is quantity measured in experiment:

$$\tau(x,y,\pm h,t) = T_s - \langle T \rangle_h. \tag{11}$$

The solution of the boundary value problem, Eqs. (10,11), is

$$\tau(\mathbf{r},t) = 0.5[(c/\lambda)\langle T_t \rangle_h - (q/\lambda) - L_{xy}\langle T \rangle_h](z^2 - h^2) + T_s - \langle T \rangle_h. \tag{12}$$

According to Eq. (9) $\langle \tau \rangle_h = 0$. From this we have

$$c = [q + \lambda L_{xy}\langle T \rangle_h + 3(\lambda/h^2)(T_s - \langle T \rangle_h)]/\langle T_t \rangle_h. \tag{13}$$

In a case of Ohmic heating $q = \sigma E^2$, where $E$ is the intensity of the electric field in a sample.
Thus measurements of $q$, $\langle T \rangle_h$, $L_{xy}\langle T \rangle_h$, $T_s$, and $\langle T_t \rangle_h$ during heating allow one to establish relation between $c$ and $\lambda$, corresponding to Eq. (13).
In a cooling stage $q = 0$ and the temperature conductivity coefficient can be expressed through measured quantities by the following relationship:

$$a = \lambda/c = h^2\langle T_t \rangle_h/[h^2 L_{xy}\langle T \rangle_h + 3(T_s - \langle T \rangle_h)]. \tag{14}$$

Calculated value of $a$ could be used to treat data, obtained during heating, and to calculate $c$ according Eqs. (13,14):

$$c = q/[\langle T_t \rangle_h - aL_{xy}\langle T \rangle_h + (3a/h^2)(\langle T \rangle_h - T_s)]. \tag{15}$$

The thermal conductivity $\lambda = ac$ could be obtained from Eqs. (14,15).
Power of the thermal losses from the unit area of the surface of a sample $P$ can be also calculated, using measured quantities. According to the definition of the thermal losses, and taking into account Eqs. (12,13) we have

$$P = -\lambda(\partial\tau/\partial z)_{z=\pm h} = 3\lambda(\langle T \rangle_h - T_s)/h. \tag{16}$$

Thus Eqs. (14-16) determine basic thermophysical characteristics in the case of a thin slab bulk heated sample.

### 3.2 Cylinder

Let us consider a cylinder (a wire) having circular section of area $s$ and radius $R$. Let $\langle T \rangle_s$ denotes the temperature averaged over the section of a cylinder. We assume that the temperature of the surface $T_s$ changes smoothly along the cylinder (coordinate $z$), i.e., $T_s$ changes inessentially on the length of $R$:

$$R \, |\partial \ln(T_s/T_0)\partial z \, | \ll 1. \tag{17}$$

According to Eq. (5)

$$T(r,z,t) = \langle T \rangle_s + \tau(r,z,t), \quad \tau(r,z,t) \ll \langle T \rangle_s. \tag{18}$$

Temperature $\langle T \rangle_s$ depends on time and changes smoothly along the axis $z$. Smooth dependence of parameters on $z$ allows one to neglect $L_z\tau$ comparatively to $L_s\tau$ ($L_s$ is a sectional part of the Laplacian in polar coordinates) in thermal conductivity equation and to obtain relation analogous to Eq. (7):

$$L_s\tau = (c/\lambda)\langle T_t \rangle_s - (q/\lambda) - L_z\langle T \rangle_s. \tag{19}$$

In this equation $z$ and $t$ acts as mere parameters, not variables. Boundary condition for Eq. (19) is

$$\tau(R,z,t) = T_s - \langle T \rangle_s. \tag{20}$$

Considered problem has axial symmetry. Due to the heat losses from the surface of a cylinder the maximum value of the temperature in a sample is on its axis. Therefore we have

$$(\partial \tau/\partial r)_{r=0} = 0. \tag{21}$$

Boundary value problem, Eqs. (19-21), has the following solution:

$$\tau = \{[(c/\lambda)\langle T_t \rangle_s - (q/\lambda) - L_z\langle T \rangle_s](r^2 - R^2)/4\} + T_s - \langle T \rangle_s. \tag{22}$$

From Eq. (18) follows that $\langle \tau \rangle_s = 0$. Eq. (22) and this relation yield:

$$c = [qR^2 + \lambda R^2 L_z\langle T \rangle_s + 8\lambda(T_s - \langle T \rangle_s)]/R^2\langle T_t \rangle_s. \tag{23}$$

During cooling stage $q = 0$ and Eq. (23) yields:

$$a = R^2\langle T_t \rangle_s /[R^2 L_z\langle T \rangle_s + 8(T_s - \langle T \rangle_s)]. \tag{24}$$

Relation analogous to Eq. (15) is as follows:

$$c = qR^2/[R^2\langle T_t \rangle_s - aR^2 L_z\langle T \rangle_s + 8a(\langle T \rangle_s - T_s)]. \tag{25}$$

The thermal conductivity $\lambda = ac$ could be obtained from Eqs. (24,25).
Power of the thermal losses from the unit area of the surface of a sample is as follows:

$$P = -\lambda(\partial \tau/\partial r)_{r=R} = 4\lambda(\langle T \rangle_s - T_s)/R. \tag{26}$$

Eq. (26) was derived taking into account Eqs. (22,23).

### 3.3. Sphere

Let us consider a sphere of a radius $R$. We assume that the temperature on the sphere $T_s(t)$ is constant all over the surface. Solution of Eq. (7), which satisfies condition $\tau(R,t) = T_s(t) - \langle T \rangle$, can be written as

$$\tau(r,t) = \{[(c/\lambda)\langle T_t \rangle - (q/\lambda)](r^2 - R^2)/6\} + T_s - \langle T \rangle. \qquad (27)$$

It follows from Eq. (5) that $\langle \tau \rangle = 0$. Averaging Eq. (27) over the volume of a sphere yields

$$c = [qR^2 + 15\lambda(T_s - \langle T \rangle)]/R^2 \langle T_t \rangle. \qquad (28)$$

In a cooling stage Eq. (28) yields

$$a = R^2 \langle T_t \rangle / 15(T_s - \langle T \rangle). \qquad (29)$$

Relation analogous to the Eq. (15) is as follows:

$$c = qR^2/[R^2 \langle T_t \rangle + 15a(\langle T \rangle - T_s)]. \qquad (30)$$

Power of the thermal losses from the unit area of the surface of a sample is as follows:

$$P = -\lambda(\partial \tau/\partial r)_{r=R} = 5\lambda(\langle T \rangle - T_s)/R. \qquad (31)$$

Eq. (31) was derived taking into account Eqs. (27,28).

### 4. Temperature distribution and specific heat capacity: slab, cylinder and sphere

It follows from Eqs. (5,12,13) that in a sample of a slab shape

$$T(z) = T_s + 1.5(\langle T \rangle_h - T_s)[1 - (z/h)^2]. \qquad (32)$$

Eqs. (5,22,23) yield for the sample of a cylindrical shape

$$T(r) = T_s + 2(\langle T \rangle_s - T_s)[1 - (r/R)^2]. \qquad (33)$$

From Eqs. (5,27,28) follows that for a sample of spherical shape

$$T(r) = T_s + 2.5(\langle T \rangle - T_s)[1 - (r/R)^2]. \qquad (34)$$

From Eqs. (32-34) follows that the maximal values of the temperature in the sections of a slab, cylinder and sphere are correspondingly:

$$T(0) = 1.5\langle T \rangle_h - 0.5T_s, \qquad (35)$$

$$T(0) = 2\langle T \rangle_s - T_s \text{ and} \qquad (36)$$

$$T(0) = 2.5\langle T \rangle - 1.5T_s. \qquad (37)$$

Maximal temperatures are always in the center of the sample. Hence in the center of the sample the temperature gradient is zero. This means that the heat, evolving in the center of the sample is spent on the temperature increase only. This allows us to immediately write the expressions for the specific heat capacity for the slab, cylinder and sphere correspondingly:

$$c = q/(1.5\langle T_t\rangle_h - 0.5T_{ts}), \tag{38}$$

$$c = q/(2\langle T_t\rangle_s - T_{ts}) \text{ and} \tag{39}$$

$$c = q/(2.5\langle T_t\rangle - 1.5T_{ts}). \tag{40}$$

Here $T_{ts} = \partial T_s/\partial t$.

## 5. High rate heating approximation

When the rate of the heating is high, the terms $L_{xy}\langle T\rangle_h$ and $L_z\langle T\rangle_s$ are small and can be neglected. Using Eqs. (13,23,28), one can obtain for the slab

$$\lambda = qh^2(\langle T_t\rangle_h - T_{ts})/3(3\langle T_t\rangle_h - T_{ts})(\langle T\rangle_h - T_s). \tag{41}$$

For the cylinder we have

$$\lambda = qR^2(\langle T_t\rangle_s - T_{ts})/8(2\langle T_t\rangle_s - T_{ts})(\langle T\rangle_s - T_s). \tag{42}$$

For the sphere we have expression valid not only for a high heating rate, but more general, valid for the arbitrary rate of heating:

$$\lambda = qR^2(\langle T_t\rangle - T_{ts})/5(5\langle T_t\rangle - 3T_{ts})(\langle T\rangle - T_s). \tag{43}$$

Using Eqs. (16,26,31,41-43), one can obtain expressions for the power of the thermal losses from the unit area of the surface of a sample. For the slab we have:

$$P = qh(\langle T_t\rangle_h - T_{ts})/(3\langle T_t\rangle_h - T_{ts}). \tag{44}$$

For the cylinder we have:

$$P = qR(\langle T_t\rangle_s - T_{ts})/2(2\langle T_t\rangle_s - T_{ts}). \tag{45}$$

For the sphere we have more general expression, valid for the arbitrary rate of heating:

$$P = qR(\langle T_t\rangle - T_{ts})/(5\langle T_t\rangle - 3T_{ts}). \tag{46}$$

Eqs (38-46) allow one calculating the values of TPC for slab and cylinder in high rate heating experiments and for sphere in arbitrary rate of heating experiment.

## 6. Thermal expansion of the inhomogeneously heated samples

A method to measure temperature, based on the measurements of the variation of the linear dimensions of a sample and translating them to temperature, was discussed above. Let us discuss it in more details.

In one-dimensional case when $T = T(z)$, thermal stresses do not arise [20], and there is a simple relation of linear dimensions of a body and temperature:

$$\delta h/h = \delta b/b = \alpha \delta T. \tag{47}$$

Hence during measuring of the thickness of the thin slab and the lattice parameter in a surface layer the following relation takes place:

$$\langle T \rangle_s = T_s + \alpha^{-1}[(\delta h/h) - (\delta b/b)_s]. \tag{48}$$

In the two- and three- dimensional cases of inhomogeneous temperature distribution in a body thermal stresses arise.

Dilatation due to inhomogeneous temperature distribution in a case of the axial distribution of the temperature in a cylindrical sample with free boundary conditions is given by the following relation [20,21]:

$$u_{ll}(r) = 3\alpha[T(r) - T_s] + 2[\langle T \rangle_s - T(r)](1 - 2\nu)/(1 + \nu), \tag{49}$$

where $\nu$ is the Poisson ratio.

Mean value of the stress tensor in a cylindrical sample with free boundary conditions is zero [21]. Using this fact, Eq. (49), $(\delta b/b)_s = u_{ll}(R)/3$, and $\delta R/R = \langle u_{ll} \rangle/3$, one can write:

$$\langle T \rangle_s = T_s + 3[(\delta R/R) - (\delta b/b)_s](1 - \nu)/\alpha(1 + \nu). \tag{50}$$

When $\delta R/R$ and $T_s$ are measured, the following equation is applicable:

$$\delta R/R = \int_{T_0}^{T_s} \alpha(\tau)d\tau + \alpha(T_s)(\langle T \rangle_s - T_s). \tag{51}$$

The first term in the right-hand part of Eq. (51) describes an expansion of a sample during heating from the temperature $T_0$ to the temperature $T_s$; the second term corresponds to the additional expansion due to the heating of the bulk of a sample (thermal stresses are taken into account). From Eq. (51) follows:

$$\langle T \rangle_s = T_s + [\delta R/R\alpha(T_s)] - [1/\alpha(T_s)] \int_{T_0}^{T_s} \alpha(\tau)d\tau. \tag{52}$$

Eq. (51) determines $\delta R/R$ as a function of $\alpha(\tau)$, $\langle T \rangle_s$, and $T_s$. As $\langle T \rangle_s$ only slightly differs from $T_s$, the second term in the right-hand part of Eq. (51) can be neglected. Thence we have

$$\alpha(T_s) = \partial(\delta R/R)/\partial T_s \tag{53}$$

with good accuracy.

Radial components of the displacements in a sphere with spherical symmetry of the distribution of the temperature is as follows [21]:

$$u = \alpha[(1 + \nu)/r^2(1 - \nu)] \int_0^r [T(\rho) - T_s]\rho^2 d\rho + [2r(1 - 2\nu)/3(1 - \nu)](\langle T \rangle - T_s). \tag{54}$$

Eq. (54) allows calculating dilatation and then $\langle T \rangle = T_s + \ldots$ Equation for $\langle T \rangle = T_s + \ldots$ coincides with Eq. (50) for a sample of a cylindrical shape.

Expressions for TPC written above contain also $\langle T_t \rangle_h$, $\langle T_t \rangle_s$, and $\langle T_t \rangle$. They can be expressed as

$$\langle T_t \rangle_h = h_t/\alpha h, \quad \langle T_t \rangle_s = \langle T_t \rangle = R_t/\alpha R \tag{55}$$

when small terms of the order of $(\langle T \rangle_h - T_s)/\langle T \rangle_h$, $(\langle T \rangle_s - T_s)/\langle T \rangle_s$, and $(\langle T \rangle - T_s)/\langle T \rangle$ are neglected.

In Eq. (55) $h_t = \partial h/\partial t$ and $R_t = \partial R/\partial t$.

In this section we have obtained equations, expressing variation of the temperature and changes in the dimensions of samples and lattice parameters in a surface layer, coefficients of the thermal expansion, and the Poisson ratio.

## 7. Discussion

As was mentioned above, the concept and the ideology of the express way of the complex measuring of TPC of conductive materials was proposed in 1987 by Meir E. Gurevich in fond memory. The author of the present paper formulated and solved theoretical problems in this topic and obtained theoretical results discussed here. Experimental, instrumental, and technological aspects of this research were performed by Meir E. Gurevich in cooperation with V. N. Minakov and his coworkers, E. N. Bludilin and A. F. Zhuravlev. Results of this research were published in [1-3]. The aim of this paper was revision and better presentation of the theoretical part of the research and introduction of these results to English language readers.

Theoretical results, described in **Sections 4** and **5** (Eqs. (32-46)) should be used in this topic to treat experimental data, independently of the available instruments and methods of measurements of averaged and surface temperatures of the samples under investigation.